\documentclass[journal]{IEEEtran}
\usepackage{graphicx}
\usepackage{booktabs,tabularx,array}
\usepackage{cite,url}
\usepackage{amsmath,amssymb,amsfonts}
\usepackage{algorithmic,textcomp,makecell}
\usepackage[table]{xcolor}
\usepackage{placeins}
\usepackage[hidelinks]{hyperref}

\begin{document}

\title{Intent-Defined Radio: A Goal-Native Architecture for Next-Generation Wireless Networks}

\author{Wencheng~Yan, ~\IEEEmembership{Student~Member,~IEEE,} 
Binrui~Xu, ~\IEEEmembership{Student~Member,~IEEE,} 
Yihao~Shen, ~\IEEEmembership{Student~Member,~IEEE,} 
Zhiyuan~Jiang, ~\IEEEmembership{Member,~IEEE,}
and Sheng~Zhou, ~\IEEEmembership{Senior~Member,~IEEE}%

\thanks{W. Yan, B. Xu, and S. Zhou (Corresponding Author) are with the Department of Electronic Engineering and the Beijing National Research Center for Information Science and Technology, Tsinghua University, Beijing, China. 

Y. Shen and Z. Jiang are with the School of Communications and Information Engineering, Shanghai University, Shanghai, China.}}%

\raggedbottom
\maketitle

\begin{abstract}
Wireless systems are becoming harder to operate because service goals are increasingly expressed in outcome-oriented terms, while radio platforms still expose mainly low-level configurations and protocol-specific controls. This mismatch makes it difficult to translate latency, reliability, throughput, and continuity requirements into coherent radio behavior. This article argues for \textbf{Intent-Defined Radio (IDR)} as a new system abstraction for next-generation wireless networks. Rather than treating the radio stack as a fixed mechanism set to be manually engineered and then tuned, IDR connects explicit service goals to executable radio behavior within a constrained design space by leveraging the capacity of large language models (LLMs). We position IDR relative to programmable radio, intent-based networking, and learning-based protocol design. We also present two simulation studies using Intent2MAC to illustrate the design loop, demonstrating executable MAC synthesis and offline refinement guided by evaluation feedback. The results show improved service satisfaction at a throughput cost and improved remote-control performance when random contention is replaced by distributed urgency-based arbitration. We finally discuss the open challenges in intent specification, evidence, hardware execution, interoperability, and lifecycle cost, and outline a staged path toward deployment.

\end{abstract}


\section{Introduction: From Radio Configuration to Intent-Defined Radio}

\begin{figure*}[t]
\centering
\includegraphics[width=0.95\textwidth]{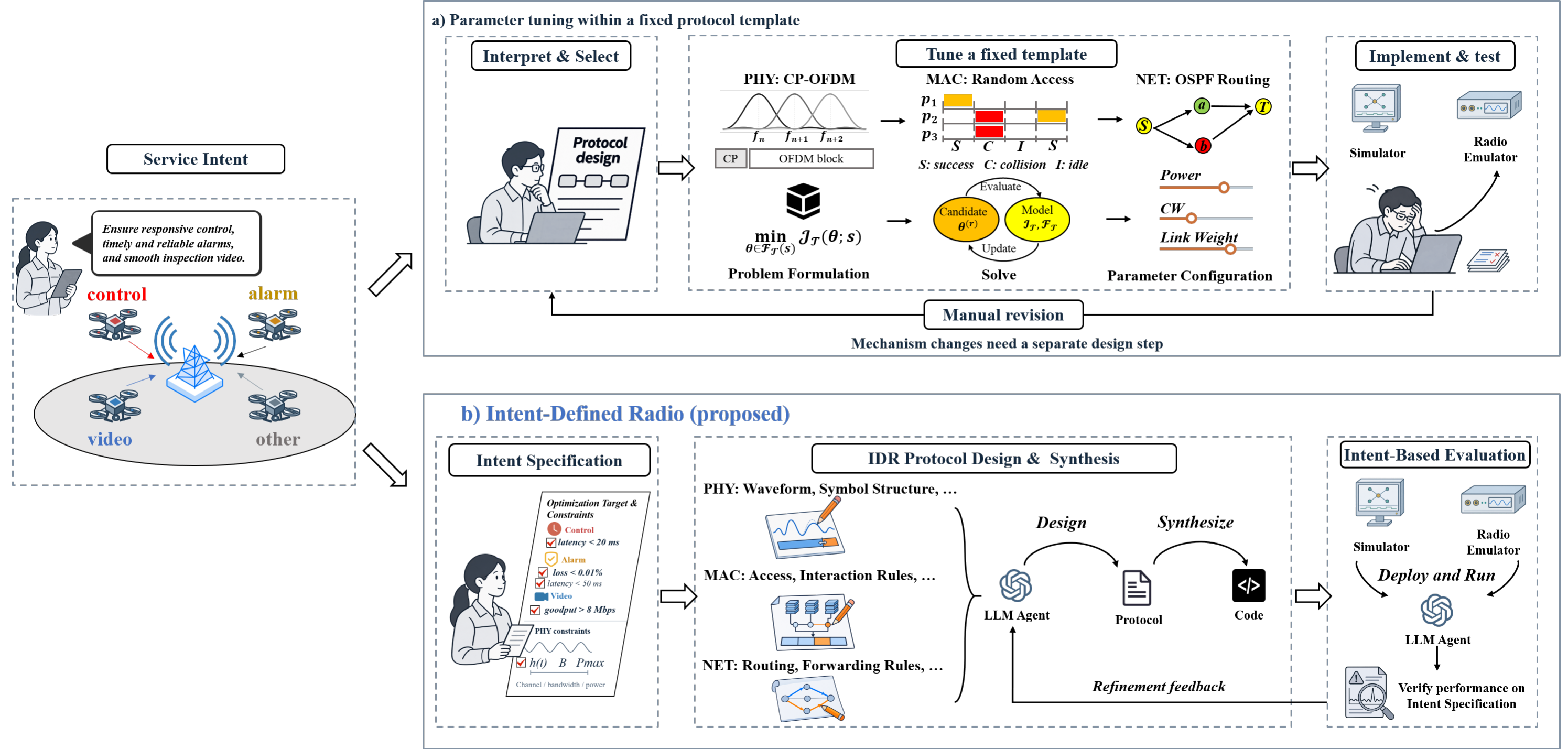}
\caption{Motivating example and conceptual transition from traditional networking toward Intent-Defined Radio (IDR). }
\label{fig:idr_example}
\end{figure*}

The diversity of wireless services is now growing faster than the protocol and control frameworks used to support it. Emerging wireless services impose heterogeneous requirements in several specific scenarios such as industrial automation, low-altitude connectivity, integrated sensing and communications, and massive low-power access. For example, unmanned aerial vehicles (UAVs) performing cooperative inspection carry control traffic, inspection video, and collected data~\cite{3gpp_uas}. Considering the illustrative inspection fleet in Fig.~\ref{fig:idr_example}, control and alarm messages need timely delivery, video needs a useful data rate, and background uploads may tolerate delay. The application specifies the desired outcomes; a radio design must determine how to realize them with limited resources. 

This translation involves choosing access procedures, queue policies, retransmission rules, and coordination timing, as well as parameter values. In a template-based workflow, an engineer selects a protocol structure and optimization adjusts settings within it. Structural changes require a separate design step. Programmable radio, cognitive radio, and AI-defined radio access networks provide foundations for reconfiguration, learning, and cross-layer optimization~\cite{mitola,haykin,yaoran}, while intent-based networking (IBN) makes desired outcomes and their assurance explicit~\cite{rfc9315}. The opportunity is to connect these capabilities in a radio-specific design lifecycle.

We propose \emph{Intent-Defined Radio (IDR)} as an architecture in which structured service intent guides the design, execution, and revision of bounded radio behavior. A behavior specifies mechanisms, interaction rules, timing, parameters, and admissible runtime changes. The central question is which changes are permitted to satisfy an intent, and what evidence justifies their use. This requires both an explicit design space and checks beyond aggregate performance.

Specifically, this article makes three contributions. First, we clarify the abstraction gap between service intent and radio mechanisms and position IDR relative to programmable and intent-based networking paradigms. Second, we present an IDR architecture that connects intent interpretation, bounded behavior synthesis, execution, verification, and evolution. Third, we present Intent2MAC as a partial realization through medium access control (MAC) synthesis. An industrial multi-robot case uses packet-level simulation; a remote-control case uses a separate process-and-communication simulator. Together they illustrate how different objectives can guide executable design revisions, while identifying the evidence still needed for operational IDR.

Sections II and III develop IDR's motivation and architecture; Section IV examines Intent2MAC, followed by practical challenges and a deployment roadmap.

\section{Why Intent-Defined Radio Matters}
Current wireless systems are not short of optimization mechanisms; they are short of a systematic abstraction that connects changing service goals to verifiable radio behavior. Most protocols are designed around relatively fixed procedures and are subsequently adapted through parameter tuning, feature profiles, or scenario-specific control logic. This approach is effective when service classes and operating conditions remain predictable, but becomes increasingly brittle as requirements vary across traffic mixes, mobility states, deployment environments, and business priorities.

\begin{itemize}
    \item \textbf{Semantic gap}. Service owners express requirements as outcomes—bounded latency, reliability, continuity, fairness, or energy efficiency—whereas radio systems expose low-level mechanisms such as scheduler priorities, retransmission timing, resource reservation, and measurement policies. Intent-based networking makes desired outcomes and their assurance explicit~\cite{rfc9315}; this motivates an inspectable mapping to radio-level mechanisms.
    \item \textbf{Design gap}. Programmability and learning enlarge the available choices, but each target platform still imposes an admissible design space. IDR should make that space explicit: which mechanisms can be composed, which timing rules can change, and which protocol invariants must remain intact. The resulting specification must be realizable on the chosen execution backend.
    \item \textbf{Assurance gap}. As wireless intelligence moves from parameter adaptation toward mechanism-level redesign, performance improvement alone is no longer sufficient. A generated radio behavior must also satisfy hard constraints, preserve protocol invariants, expose its decision path, and support rollback or fallback when the active intent is violated. This motivates assurance methods that go beyond empirical performance alone~\cite{seshia}.
\end{itemize}

\begin{table*}[t]
\centering
\caption{Related approaches and their relationship to the proposed IDR architecture.}
\label{tab:paradigm_comparison}
\small
\setlength{\tabcolsep}{5pt}
\renewcommand{\arraystretch}{1.15}
\begin{tabularx}{\textwidth}{
>{\hsize=0.85\hsize\linewidth=\hsize\centering\arraybackslash}X
>{\hsize=0.85\hsize\linewidth=\hsize\centering\arraybackslash}X
>{\hsize=1.15\hsize\linewidth=\hsize\centering\arraybackslash}X
>{\hsize=1.15\hsize\linewidth=\hsize\centering\arraybackslash}X}
\toprule
\textbf{Approach} & \textbf{Primary abstraction} & \textbf{Typical scope} & \textbf{Relationship to IDR} \\
\midrule
Software-defined radio~\cite{mitola} & Software programmability & Waveforms, protocol functions, and parameters & Provides an execution substrate; does not prescribe an intent model \\
\midrule
Cognitive radio~\cite{haykin} & Environment-aware learning & Radio decisions informed by sensing and experience & Supplies adaptation principles; complements explicit service intent \\
\midrule
O-RAN~\cite{poleseoran} & Open RAN interfaces & Control applications, policies, and configuration & Supplies control and execution interfaces with defined scope \\
\midrule
IBN~\cite{rfc9315} & Desired service outcomes & Intent translation, fulfillment, and assurance & Supplies intent semantics; radio synthesis is an implementation choice \\
\midrule
Learning-based protocol design~\cite{macsynthesis,llm4mac,cpagentnet} & Learned policies or generated mechanisms & Signaling, access, mechanism composition, and strategy generation & Supplies synthesis methods that can be integrated into the lifecycle \\
\midrule
\textbf{IDR (proposed)} & \textbf{Intent, constraints, and evidence} & \textbf{Bounded behavior design and its lifecycle} & \textbf{Connects synthesis to deployment, adaptation, and future design updates} \\
\bottomrule
\end{tabularx}
\par\smallskip
\end{table*}

Table~\ref{tab:paradigm_comparison} positions these interfaces relative to existing approaches. O-RAN supplies control interfaces~\cite{poleseoran}; distributed learning and generative agents already support protocol design~\cite{macsynthesis,llm4mac,cpagentnet}. IDR's proposed contribution is to carry an explicit radio-design specification and its supporting evidence across synthesis, execution, and revision. Its scope can expand from configuration to bounded mechanism redesign as platform capabilities and validation permit. The following 4S architecture organizes that lifecycle.

\section{An Architecture for Intent-Defined Radio}

\begin{figure*}[t]
\centering
\includegraphics[width=1.0\textwidth]{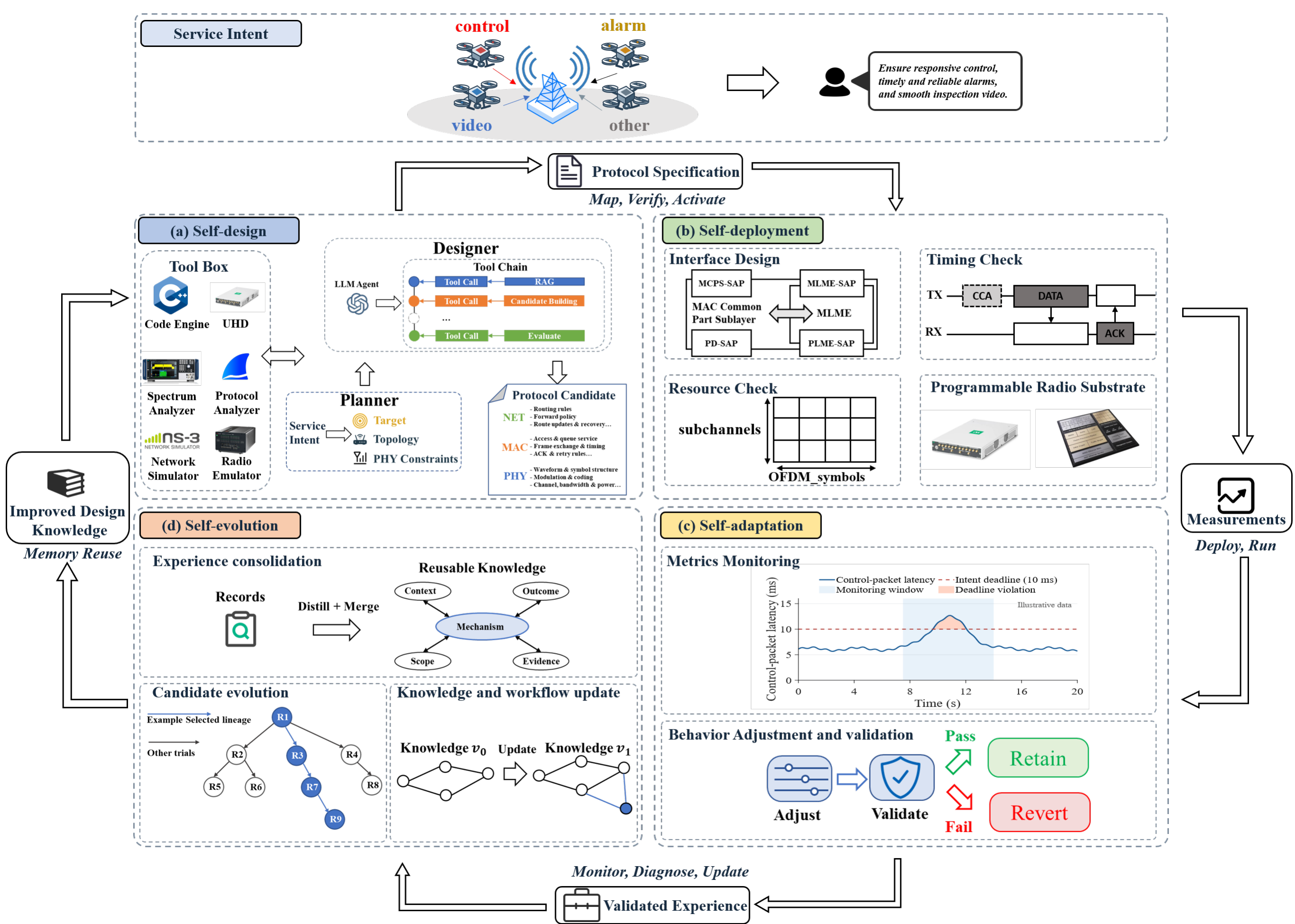}
\caption{Proposed IDR architecture.}
\label{fig:architecture}
\end{figure*}

Fig.~\ref{fig:architecture} organizes IDR around service intent and four connected capabilities. Intent specifies desired outcomes, priorities, and constraints; observed results determine whether those outcomes are achieved. The loop links protocol content, its implementation, current operating behavior, and knowledge retained for later versions.

Learning-based MAC synthesis already explores distributed reinforcement learning~\cite{macsynthesis}, while LLM-based approaches explore learned and agent-driven protocol decisions~\cite{llm4mac,cpagentnet}. IDR's architectural distinction is an explicit protocol specification linking synthesis to platform adaptation, runtime control, and subsequent revision. LLM reasoning can guide design and learning, while deployed mechanisms execute through radio interfaces on the required timescale. Realization therefore requires platform adapters, monitoring, and feedback validation alongside the synthesis engine.

\noindent\textbf{Self-design.}
The planner translates service intent into an explicit objective, scenario assumptions, and platform constraints. Using these inputs and reusable design knowledge, the LLM-based designer calls tools to retrieve relevant knowledge, build candidates, and evaluate their behavior. Feedback guides revisions to mechanisms, interaction rules, timing, and parameters within the permitted design space. The resulting protocol specification records these choices and their operating limits, providing a concrete input to self-deployment.

\noindent\textbf{Self-deployment.}
The specification is adapted to software, firmware, or reconfigurable hardware~\cite{openradio}. The figure's interface, timing, and resource blocks capture the key mappings: protocol functions to device data and management interfaces, sensing and message exchanges to hardware timing, and processing to available resources. Functional and timing verification precedes coordinated activation on participating devices.

\noindent\textbf{Self-adaptation.}
When channel conditions or traffic change, monitoring relates link measurements to service outcomes. Runtime control then adjusts deployed behavior, for example by selecting more robust transmissions or redistributing access opportunities. Feedback validation checks whether service recovers within the protocol's operating limits. Persistent violations trigger a new design and deployment cycle. Fast packet processing remains separate from slower design reasoning.

\noindent\textbf{Self-evolution.}
Experience from multiple runs records which mechanisms succeed or fail and under what conditions. As shown on the left, experience consolidation extracts reusable lessons, candidate evolution compares revised protocols, and capability updates improve design knowledge, templates, or the synthesis agent. Better subsequent versions should retain earlier strengths and improve performance under new conditions. This extends autonomic learning~\cite{kephart} toward improving future radio designs.

The cases in Section~IV demonstrate protocol generation and refinement in simulation. Hardware deployment, validated runtime adaptation, and sustained improvement across deployments remain goals of the proposed architecture.

\section{Representative Cases: Intent-Driven MAC Synthesis}
\label{sec:intent2mac_case}

\begin{figure*}[t]
\centering
\includegraphics[width=0.95\textwidth]{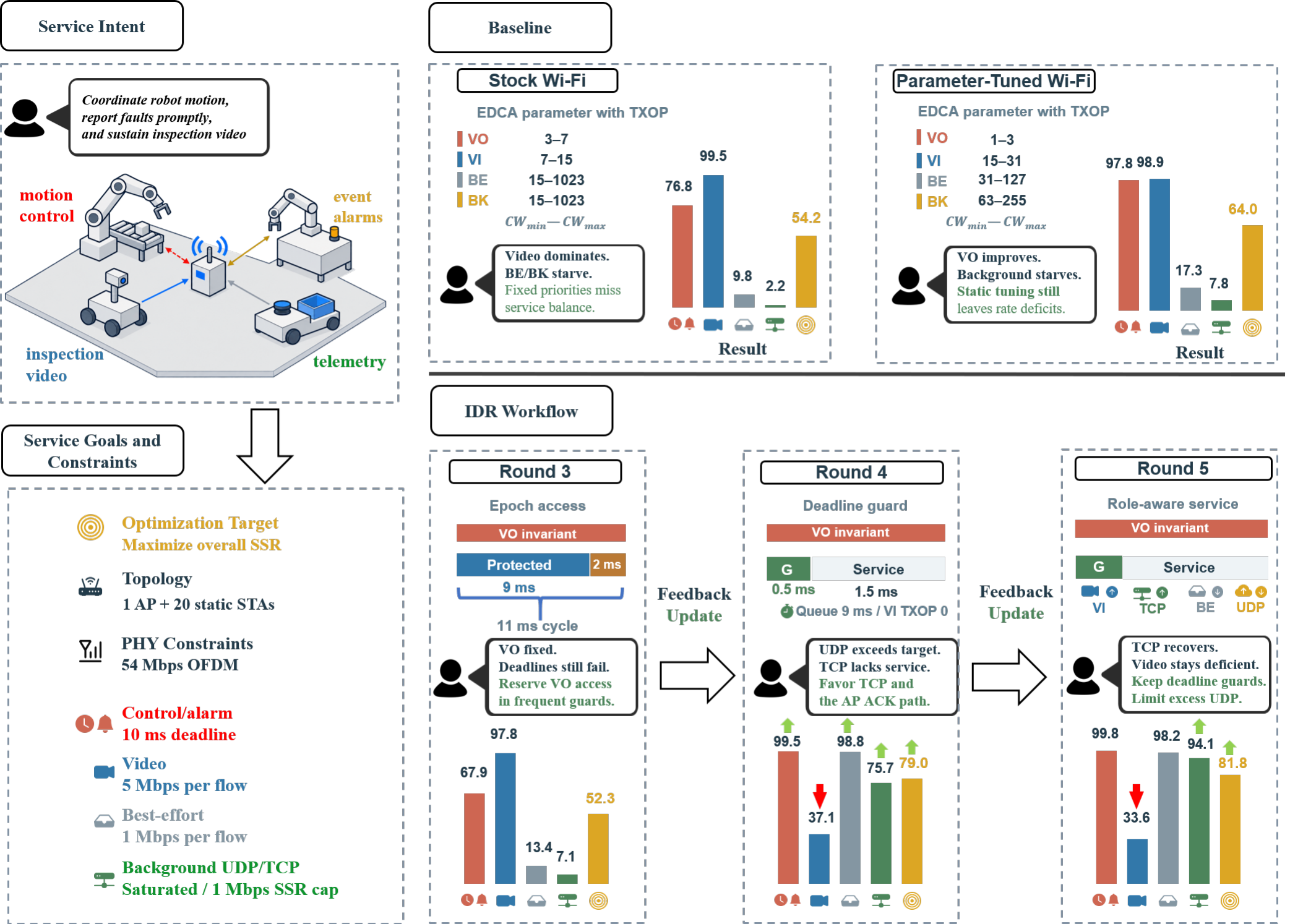}
\caption{Intent2MAC for industrial multi-robot coordination: service intent, Wi-Fi baselines, and feedback-guided MAC refinement. Bars show grouped service satisfaction and overall SSR.}
\label{fig:intent2mac_case}
\end{figure*}

Both studies used OpenAI GPT-5.6 as the Intent2MAC design agent. Humans supplied the task and execution boundaries without participating in protocol design. Inputs comprised service goals, frozen scenarios, platform interfaces, reference implementations, and permitted changes. Within each round, the agent formalized the requirement, produced or revised the protocol specification, and translated it into a JSON domain-specific language (DSL) and backend code with implementation bindings. The DSL records mechanisms, states, timers, dependencies, and trace predicates. The ns-3 study binds actions to Wi-Fi MAC hooks in C++, whereas Python/Gym uses schema-checked rules and a timed distributed-MAC executor. These are separate implementation paths, without a shared automatic DSL compiler.

Tools then built or validated the implementation, ran the simulation, and collected performance metrics and mechanism traces. Candidate validity required successful backend checks and execution, evidence that the intended mechanisms operated, and satisfaction of any explicitly specified hard constraints. Feedback grounded in these results guided revisions to the protocol specification before the next round’s DSL and code were generated. Both studies budgeted five complete feedback-linked rounds and retained valid low-performing candidates. Implementation failures were repaired within the same round, with up to five failed attempts permitted per round. The candidate with the highest mean primary-objective score among valid candidates was selected, yielding R5 in both studies. During simulation, fixed MAC rules executed without per-packet LLM decisions.

\subsection{Intent-Guided Service Balancing in Industrial Robot Networks}

Industrial robot teams share wireless links for motion coordination, fault reporting, visual inspection, and routine telemetry. Late control or alarms can interrupt coordinated work, video requires sustained delivery, and background transfers can tolerate waiting. The service intent is to maximize task completion across the team; the evaluation represents this intent through communication-level service satisfaction rather than measured robot operations.

Figure~\ref{fig:intent2mac_case} uses an ns-3 model with 20 static stations sharing one AP over a 54-Mb/s link. Service satisfaction ratio (SSR) averages twenty per-flow scores equally: seven control/alarm flows (five control and two alarm), five video, four best-effort, and four background flows (two TCP and two UDP). Each grouped bar averages its member-flow scores. Control and alarm satisfaction is the fraction of generated packets received within 10~ms; rate-based satisfaction is goodput divided by the target, capped at one. Targets are 5~Mb/s per video flow and 1~Mb/s per best-effort or background flow. All flows are weighted equally, and the objective permits tradeoffs among services without requiring every target to be met. Topology, PHY, and traffic are shared across comparisons.

The baselines are stock Wi-Fi and the best of twelve predefined static enhanced distributed channel access (EDCA) profiles. Static tuning improves urgent service but leaves best-effort and background flows underserved. IDR Candidate R3 retained long access epochs and left urgent deadlines unmet despite eliminating runtime VO-profile writes. Its feedback therefore called for short, frequent deadline guards, introduced in R4. Control/alarm satisfaction increased from 67.9 to 99.5 percent, while overall SSR rose from 52.3 to 79.0 percent.

IDR Candidate R5 combines a recurring deadline guard with role-aware contention. Each 2-ms microcycle contains a 0.5-ms guard and a 1.5-ms service phase. During the guard, nonurgent traffic receives large contention windows and longer access deferral, favoring control and alarms in the voice (VO) access category. VO parameters remain fixed during phase changes, and its queue lifetime is capped at 9~ms to limit service of aging packets. Expired packets remain in the satisfaction denominator. Protection remains contention based: transmissions already in flight finish normally, and deadlines are not guaranteed. Standard Wi-Fi framing, acknowledgment, and retry procedures are retained.

R4 feedback identified excess UDP service above the capped target while TCP remained underserved. R5 therefore retained the guards and replaced uniform service-phase contention profiles with role-specific profiles, including for devices sharing the same access category. Video and the AP's returning TCP acknowledgments receive aggressive access. Relative to R4, background TCP satisfaction rises from 51.3 to 88.2 percent, while the combined TCP/UDP background score increases from 75.7 to 94.1 percent. Application traffic generators and transport protocols remain unchanged.

Averaged across three matched simulation seeds, IDR Candidate R5 reaches 81.8 percent SSR, versus 54.2 percent for stock Wi-Fi and 64.0 percent for static tuning. Recorded phase transitions and role-profile writes confirm that the intended mechanisms execute.

The gain comes with aggregate throughput falling from 26.7 to 18.3~Mb/s relative to stock, and video attainment remaining at 33.6 percent. The objective rewards recovering several underserved flows even when video loses rate; a mandatory video target would reject this candidate. 

\subsection{Context-Aware Status Updates for Remote Control}

\begin{figure*}[t]
\centering
\includegraphics[width=0.95\textwidth]{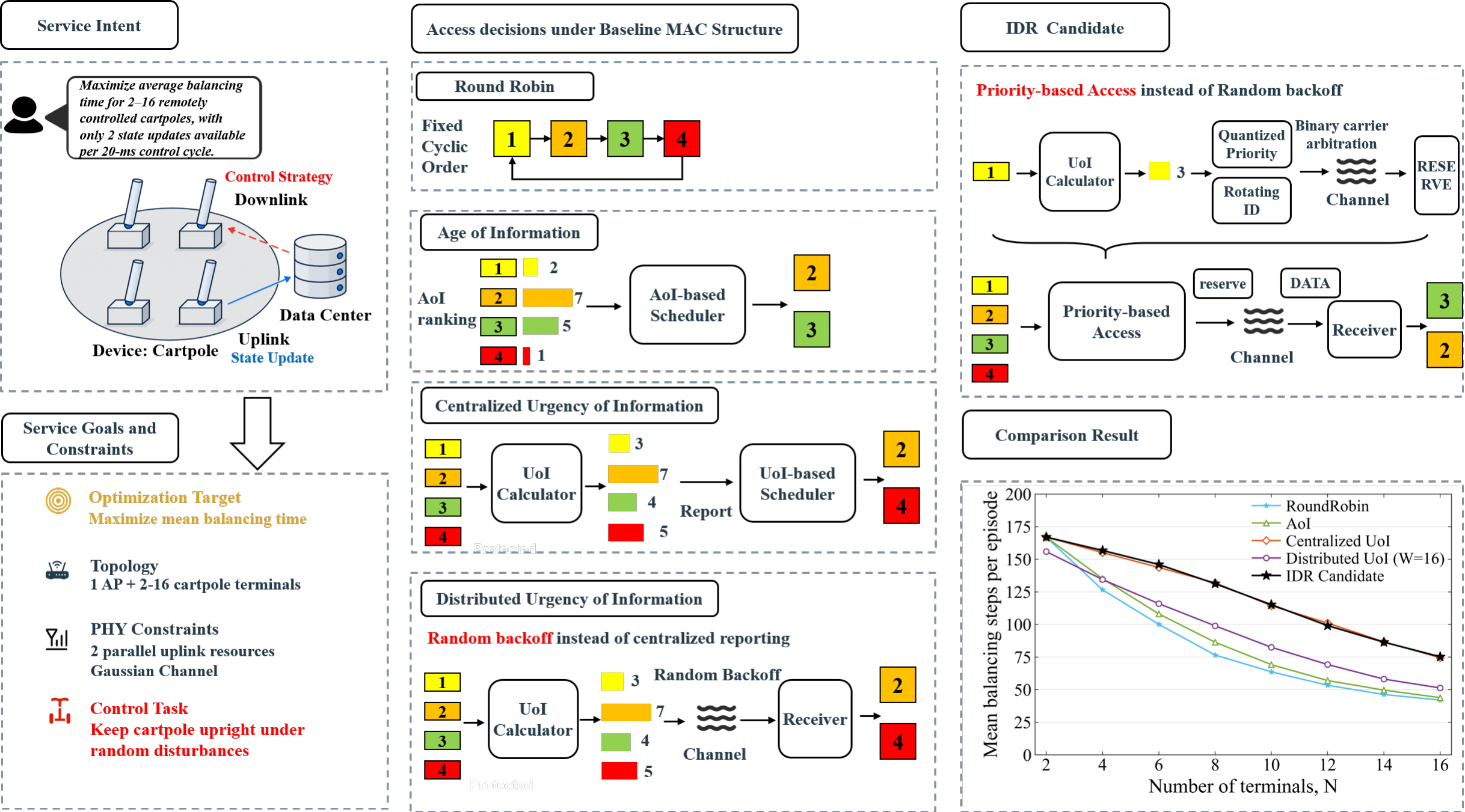}
\caption{Context-aware remote control: service intent, baseline access rules, distributed priority access, and mean balancing duration under a common communication budget.}
\label{fig:intent2mac_cartpole}
\end{figure*}

State updates of the same age can have different consequences for a pole approaching a fall. Urgency of information (UoI) captures this context by weighting estimation error according to task importance~\cite{uoi}. In Fig.~\ref{fig:intent2mac_cartpole}, a controlled Python/Gym reconstruction connects one AP to 2--16 CartPoles sharing two uplink resources per 20-ms cycle. The objective is mean balancing duration under random disturbances, capped at 200 steps per episode, including failures and equally weighting terminals and terminal-count settings. All schemes share the controller, dynamics, signaling budget, and next-boundary availability of received updates.

Four baselines represent different access principles. Round robin serves terminals cyclically; age-of-information (AoI) scheduling prioritizes stale measurements using receiver-maintained ages and link reliability. Centralized UoI ranks reported urgency indices and grants the two highest-ranked terminals access, paying for reports and grants. Distributed UoI lets terminals above a local urgency threshold contend through random backoff and reservation within a 16 mini-slot contention window, with possible collisions or unused resources. All schemes share a 2-ms pre-DATA control phase comprising twenty 100-$\mu$s mini-slots. The distributed baseline uses a 16-slot backoff-and-reservation window, leaving the final four slots idle, whereas IDR Candidate R5 uses all twenty; DATA always starts at 2 ms, with ACK/NACK airtime accounted for separately after DATA.

Development feedback showed that random reservation still missed high-urgency updates after R3’s admission refinements. R4 therefore replaced random contention with binary priority arbitration. Its coarse urgency quantization and fixed-ID tie preference then motivated R5’s finer urgency encoding and rotating tie order. Mean balancing duration on the development runs was 96.1, 119.2, and 120.3 steps for R3, R4, and R5, respectively.

R5 retains local UoI evaluation and replaces thresholded random contention with distributed binary priority arbitration. All terminals contend using a nine-bit key: five bits logarithmically quantize urgency, and four bits encode a rotating terminal identifier. Higher urgency levels take precedence; the public clock rotates tie-breaking order within a quantized level. Quantization bounds arbitration time while preserving a coarse ordering of control relevance.

For each resource, contenders examine the key from its most significant bit. A contender whose current bit is one transmits a dominant control burst; a contender whose bit is zero listens and withdraws from that resource's arbitration if it senses a busy channel. Each survivor broadcasts a reservation. The first winner is excluded from arbitration for the second resource, where the remaining terminals repeat the process. Two nine-bit arbitrations, each followed by one reservation slot, occupy the 20 paid control mini-slots within 2~ms. The winners transmit on separate uplink resources. The AP receives and acknowledges updates without collecting priority reports or issuing access grants.

At 16 terminals, the DATA-collision fraction averages 15.3 percent for distributed UoI and zero for IDR candidate in the recorded runs. Dominant control bursts can still overlap: their busy energy is deliberately used for arbitration. These overlaps are distinct from DATA collisions.

IDR candidate was selected and frozen before the paired retest, which used seeds separate from those used for design, debugging, and candidate selection. The retest results were not fed back into protocol revision or candidate selection. Across 32 paired retest seeds and equally weighted terminal-count settings, IDR candidate achieves 122.1 balancing steps, versus 95.8 for distributed UoI and 121.8 for centralized UoI. The gain over the distributed reference is 27.5 percent. The paired difference from centralized UoI has a 95 percent confidence interval from -0.64 to 1.26 steps, providing no evidence of superiority.

Distributed coordination nevertheless incurs control cost. At 16 terminals, summed control-transmission time is 13.5 percent of elapsed time for IDR candidate, versus 10.0 percent centrally; simultaneous bursts contribute separately to this sum. The result therefore demonstrates a structural access change under a fixed communication budget. Its benefit depends on synchronized arbitration and reliable carrier observations; hidden terminals and sensing errors remain outside the model.

\subsection{Lessons Learned}

The two cases show why a service objective should guide mechanism selection as well as parameter tuning. Their benefits nevertheless concern complete, fixed candidates. The comparisons do not isolate individual mechanisms' causal contributions, and simulation seeds do not establish repeatability across independent LLM design searches. The distinct backends also need separate validation; a common workflow cannot establish equivalence between their communication models.

Relative to the 4S lifecycle, the evidence supports bounded self-design, executable instantiation in simulation, and offline refinement from execution feedback. Hardware deployment, validated runtime adaptation, and cross-task self-evolution remain to be demonstrated. Independent design searches, mechanism ablations, and timing-correct hardware experiments would help determine how reliably these gains can be reproduced and transferred.

\section{Challenges and Roadmap Towards Practical IDR}

We list several key challenges toward practical IDR implementation. Practical IDR must match generated mechanisms to hardware timing and resource limits, establish credible benefits beyond simulation, and introduce updates without losing interoperability or operator control. A higher aggregate score must also be distinguished from compliance with essential service requirements. Lifecycle cost determines whether repeated synthesis and reconfiguration are worthwhile. These challenges motivate four maturity stages in which design freedom expands with execution and validation evidence.

\noindent\textbf{Stage 1: Intent-aware adaptation within existing mechanisms.}
Begin with intent-guided selection and tuning through established interfaces. Make service priorities, hard constraints, and acceptable tradeoffs explicit before optimization, then measure how configuration changes affect the requested outcomes. Operator-approved action limits, monitored baselines, and recovery procedures provide an initial operating boundary.

\noindent\textbf{Stage 2: Bounded self-design and executable synthesis.}
Extend from profile selection to mechanism composition within a target platform's capabilities~\cite{openradio}. Protocol specifications and executable representations should expose what changes and how it is implemented. Acceptance requires working code, evidence that the intended mechanisms execute, reproducible baseline comparisons, and explicit reporting of unmet requirements. Intent2MAC illustrates this stage in simulation; repeatability across independent design searches remains a further test.

\noindent\textbf{Stage 3: Verifiable self-deployment and runtime adaptation.}
Validate execution time, memory use, and radio-interface compatibility on hardware before limited field trials~\cite{poleseoran,openradio}. Trials should examine interference, traffic changes, coexistence with unmodified devices, and disruption during coordinated updates. Runtime adaptation stays within the validated scope, with monitoring and fallback when service deteriorates. This stage connects simulated mechanisms to operational constraints.

\noindent\textbf{Stage 4: Evidence-driven and governed self-evolution.}
Use experience across deployments to improve reusable mechanisms and future synthesis~\cite{kephart,seshia}. Compare versioned revisions under renewed validation and retain deployment evidence linking each change to its benefit or failure. Assess generation, validation, reconfiguration, and maintenance costs so that better designs yield sustained operational value.

Operator authorization, interoperability, recovery, and cost assessment apply throughout. These maturity milestones can advance at different rates across the 4S capabilities.

\section{Conclusion}
This article introduced Intent-Defined Radio (IDR) as a systems abstraction for bridging service-level intent and executable radio behavior. Rather than treating wireless intelligence as a collection of isolated optimizers operating over fixed protocol structures, IDR organizes radio operation around an explicit intent-to-radio lifecycle: structured intent is translated into bounded behavior packages, verified before and during execution, and refined using runtime evidence. The resulting framework spans self-design, self-deployment, self-adaptation, and self-evolution, while keeping the chain from intent to mechanism to outcome explicit and inspectable.

The representative Intent2MAC case studies illustrate how selected elements of this vision can already be instantiated at the MAC layer, where different service intents lead to different executable mechanism structures and network trade-offs. At the same time, substantial challenges remain in intent semantics and arbitration, synthesis assurance, observability and attribution, design complexity, and practical deployment and interoperability. IDR therefore does not require unrestricted autonomy from the outset. A more credible path is to progressively expand radio design freedom together with the verification, monitoring, deployment, and recovery capabilities needed to make that freedom trustworthy.

\begin{IEEEbiographynophoto}
{Wencheng Yan} (yanwc21@mails.tsinghua.edu.cn) is a Ph.D. student in the Department of Electronic Engineering, Tsinghua University. His research interests include wireless communications, network protocol design, and edge intelligence.
\end{IEEEbiographynophoto}
\vspace{-.5in}

\begin{IEEEbiographynophoto}
{Binrui Xu} (xbr25@mails.tsinghua.edu.cn) is a master student in the Department of Electronic Engineering, Tsinghua University. His research interests include wireless communications, edge AI computing and agentic intelligent systems.
\end{IEEEbiographynophoto}
\vspace{-.5in}

\begin{IEEEbiographynophoto}
{Yihao Shen} (shenyihao@shu.edu.cn) is a Ph.D. student in the School of Communication and Information Engineering, Shanghai University. His research interests include wireless communication systems and agentic software engineering for domain-specific architectures.
\end{IEEEbiographynophoto}
\vspace{-.5in}

\begin{IEEEbiographynophoto}
{Zhiyuan Jiang} (jiangzhiyuan@shu.edu.cn) is a professor in the School of Communication and Information Engineering, Shanghai University. His research interests include wireless systems, RISC-V domain specific architectures for baseband processing, and edge AI computing.
\end{IEEEbiographynophoto}
\vspace{-.5in}

\begin{IEEEbiographynophoto}
{Sheng Zhou} (sheng.zhou@tsinghua.edu.cn) is an associate professor in the Department of Electronic Engineering, Tsinghua University. His research interests include vehicular networks, mobile edge computing, and green wireless communications.
\end{IEEEbiographynophoto}

\end{document}